\documentstyle[amssymb,prb,aps]{revtex}

\begin{document}

\twocolumn[\hsize\textwidth\columnwidth\hsize\csname
@twocolumnfalse\endcsname

\title{Anisotropic superconducting properties of aligned MgB$_{2}$ crystallites}

\author{O. F. de Lima, R. A. Ribeiro, M. A. Avila, C. A. Cardoso and A. A. Coelho}

\address{Instituto de F\'{i}sica ``Gleb Wataghin'', UNICAMP, 13083-970, Campinas, SP,
Brazil.}

\date{Mar 14, 2001}
\maketitle

\begin{abstract}
Samples of aligned MgB$_{2}$ crystallites have been prepared, allowing for the first time the direct
identification of an upper critical field anisotropy $H_{c2}^{ab}/H_{c2}^{c}=\xi _{ab}/\xi _{c}\simeq 1.7$;\ with $\xi
_{o,ab}\simeq 70$ \AA , $\xi _{o,c}\simeq 40$ \AA , and a mass anisotropy ratio $m_{ab}/m_{c}\simeq 0.3$.\ 
 A ferromagnetic background signal was identified, possibly related to the raw materials purity.


\end{abstract}

\vskip1pc] \narrowtext

The recent discovery of superconductivity at 39 K in Magnesium Diboride (MgB$%
_{2}$)\cite{Akimi} has brought new excitement to the area of basic and
applied research on superconducting materials. The observation of an isotope
effect\cite{budko}, a BCS-type energy gap measured by Scanning Tunneling
Spectroscopy\cite{anl1}, as well as band structure studies\cite
{kortus,freeman}, point to a phonon-mediated superconductivity in MgB$_{2}$.
Some reports\cite{larbal,kang} have suggested that MgB$_{2}$ has an
isotropic (or 3D) behavior, based on measurements done in polycrystalline
samples. However, other studies\cite{junod,an} have also discussed its
possible anisotropic nature. The relatively high values reported for the
critical current density\cite{larbal,canfild} ($J_{c}$) are possibly
indicating the absence of weak link problems, which are well known in the
high-T$_{c}$ materials. While polycrystalline MgB$_{2}$\ is very easy to
grow and is a readily available reagent, good-sized single crystals of this
material have not yet been reported, and their development promises to be a
greater challenge. Here we present results from samples of aligned MgB$_{2}$
crystallites that establish the anisotropy of the upper critical field ($%
H_{c2}$), thus implying an anisotropic character for other superconducting
properties, e.g., the energy gap, coherence length ($\xi $), field
penetration depth ($\lambda $), and $J_{c}.$

In this work, a weakly sintered sample of MgB$_{2}$ was prepared, starting
with a stoichiometric mixture of 99.5 at\% pure Boron and 99.8 at\% pure
Magnesium, both in chips form (Johnson Matthey Electronics). The loose
mixture was sealed in a Ta tube under Ar atmosphere, which was then
encapsulated in a quartz ampoule and put into the furnace. The compound
formation was processed by initially holding the furnace temperature at 1200$%
^{\circ }$C for 1 hour, followed by a decrease to 700$^{\circ }$C\ (10$%
^{\circ }$C/h), then to 600$^{\circ }$C\ (2$^{\circ }$C/h), and finally to
room temperature at a rate of 100$^{\circ }$C/h. The weakly sintered product
was easily crushed and milled employing mortar and pestle. Using a
stereomicroscope we could observe a very uniform powder consisting mainly of
shiny crystallites, with aspect ratios ranging from 2 to 5. This is mainly
due to the main surface size distribution ranging from 5 to 40 $\mu $m for
the larger linear dimension, since the crystallites' thickness is very
regular, around 2 $\mu $m. The powder was then sieved into a range of
particle sizes between 5 - 20 $\mu $m, which allows the crystallites
fraction to be maximized to almost 100\%. Small amounts of the powder were
then patiently spread on both sides of a small piece of paper, producing an
almost perfect alignment of the crystallites, as shown in the SEM picture in
the upper part of Fig. 1. \ The lower part of this figure shows an X-ray
diffraction pattern ($\theta -2\theta $ scan) from a sample of the\ {\it %
crystallite-painted} paper, displaying only the (001) and (002) reflections
coming from the MgB$_{2}$ phase. A lattice parameter c = 3.518 $\pm $ 0.008
\AA\ was evaluated from these two peak positions. The two small impurity
peaks marked with asterisks were indexed as SiO$_{2}$. The inset of Fig. 1
shows a rocking curve ($\omega $ scan) for the (002) peak that reveals an
angular spread around 4.6 degrees, associated with a small misalignment of
the crystallites {\bf c} axis.

Electron microprobe analysis done on four different areas between the MgB$%
_{2}$ crystallites, revealed the following average concentration (in at\%)
of elements: O (62.9), C (22.2), Ca (9.48), Si (1.48), Mg (1.44), Al (1.37),
K (0.09), Fe (0.50), Cr (0.21), Ni (0.09). The first eight elements in this
list were found also in the composition analysis made on the same type of
paper used (Canson, ref. 4567-114). Microprobe analysis done also on the
initial Mg and B revealed a few small precipitates, smaller than 10 $\mu $m
and containing up to 8 at\% Fe, only in the Mg chips. This confirms the
expectation of Fe being a common impurity\cite{raynor} in commercial Mg and
sets a general concern on its possible effects. \ The average composition
found on top of several crystallites, normalized to the whole MgB$_{2}$
formula unit,\ was: Mg (30.80), O (2.20), Ca (0.17), Si (0.07), Fe (0.06). \
Although Boron contributes with a fraction of 66.6 at\% it does not show-up
in the microprobe analysis because it is too light. The contaminants found
on top of the crystallites most possibly came from a surface contamination
caused by the alignment technique, which required vigorous rubbing on top of
the powder, using a steel tweezers tip to spread the crystallites uniformly.
This is corroborated by a further analysis done on top of several as-grown
crystallites, which detected only Mg and a small amount of O (possibly from
MgO). This result is consistent with the very small solid solubility limit
of about 0.004 at\% Fe in Mg, which is known to occur\cite{massa} at the
solidification temperature of 650 $^{\circ }$C. The inter-crystallite type
of {\it rubbish} shown in Fig. 1 is attributed mainly to the paper abrasion,
which produces a varied distribution of irregular grains of paper fragments.
\ In order to characterize the superconducting and magnetic properties of
the aligned crystallites, we mounted several samples consisting of a pile of
5 small squares (3 $\times $ 3 mm$^{2}$) cut from the {\it %
crystallite-painted} paper and glued with Araldite resin. \ Each one of
these samples contains a number of crystallites estimated to be around 6.5 $%
\times $ 10$^{5}$, totalizing an effective volume of 0.065 mm$^{3}$, which
is reasonably close to 0.060 mm$^{3}$ that was evaluated from the expected
slope of $-1/4\pi $ for the diamagnetic shielding at $H\approx 0.$

Figure 2 shows the anisotropic signature of the $H_{c2}$($T$) line in the
field interval \ $0\leq H\leq 40$ kOe. The values were taken from the
transition onset of the real component ($\chi ^{\prime }$) of ac
susceptibility, measured using a PPMS-9T machine (Quantum Design), with an
excitation field of amplitude 1 Oe and frequency 5 kHz. The inset shows an
enlarged view of the $\chi ^{\prime }$($T$) curves for H parallel (solid
symbols) and perpendicular (open symbols) to the sample {\bf c} axis. The $%
\chi ^{\prime }$($T$) as well as the $M$($T$) (inset of Fig. 3)
measurements, for $H=10$ Oe, show sharp transitions at the same critical
temperature $T_{c}=39.2$ K. The dashed lines connecting points in Figs. 2 -
4 are only guides to the eyes. Typically, some of the published data on the
temperature dependence\cite{budko,canfild,finemore,muller} of $H_{c2}$($T$)
agree with our result for $H_{c2}$($T$) // $ab$. As an example, the data
from Ref. 14 is plotted in Fig. 2 as stars. This could simply mean that in
polycrystalline samples the transitions are broadened, showing the onset at
the highest temperature that corresponds to the highest critical field
available, which is $H_{c2}(T)//ab.$

The ratio $\eta =H_{c2}^{ab}/H_{c2}^{c},$ between the upper critical field
when $H$ is applied parallel to the $ab$ plane, and when it is along the 
{\bf c }direction, was evaluated at different temperatures, producing $\eta
=1.73\pm 0.03.$ Using the Ginzburg-Landau mean field expression\cite{tinkh}
(in CGS units) \ $\xi $($T$) = $\xi _{o}\,(1-T/T_{c})^{-1/2}$ and the
results for anisotropic situations\cite{bulkii,blatt} $H_{c2}^{c}$($T$) = $%
\phi _{o}/(2\pi \,\xi _{ab}^{2})$ and $H_{c2}^{ab}/H_{c2}^{c}=1/\varepsilon $%
, where $\phi _{o}=2.07\times 10^{-7}$ G cm$^{2}$ is the quantum of flux and 
$\varepsilon ^{2}=m_{ab}/m_{c}$ is the mass anisotropy ratio, we find that $%
\xi _{o,ab}/\xi _{o,c}=\xi _{ab}$($T$)/$\xi _{c}$($T$) = $\eta \simeq 1.73$
and $\varepsilon ^{2}\simeq 0.3.$ Since at $T=27$ K we have $H_{c2}^{c}=20$
kOe, this implies that $\xi _{o,ab}\simeq 70$ \AA\ and $\xi _{o,c}\simeq 40$
\AA . \ The mass anisotropy ratio of MgB$_{2}$ thus corresponds to a
relatively small anisotropy when compared to the highly anisotropic high-Tc
cuprates\cite{blatt}, like YBCO ($\varepsilon ^{2}\simeq 0.04$) and BSCCO ($%
\varepsilon ^{2}\simeq 10^{-4}$). \ We do not expect that a likely very
small\ bulk contamination of the crystallites\ could eventually change their
anisotropy values. In fact our careful composition analysis have indicated
that almost all contaminants are located in the region between the
crystallites, thus having a negligible chance to affect the underlying
mechanism of the superconducting condensation.

The magnetization curves $M$($T$) and $M$($H$), displayed in Figs. 3 and 4,
were measured using a SQUID magnetometer (Quantum Design, model MPMS-5). The 
$M$($H$) curves ($T=5$ K) shown in Fig. 3 are intriguing in the region -1 $%
\lesssim H\lesssim $ 1 kOe, where the maximum shielding and first field
penetration (in the initial virgin state) occur. For $\left| H\right|
\gtrsim 1$ kOe the hysteretic curves in both field directions look very
similar. However, for $\left| H\right| \gtrsim 40$ kOe (not shown here) the
magnetization difference between the up and down curves ($\Delta M)$ becomes
smaller than the noise. Large fluctuations of the magnetic moment were
consistently observed in this field region, for 3 different samples and
temperatures ($T=5,10,20K$), possibly associated with the high creep rate
and the fast drop of $J_{c}$ occurring at high fields\cite
{canfild,finemore,bug,wen}. Fig. 4 shows a clear signature of the
ferromagnetic hysteresis loop measured at $T=45$ K, mainly attributed to the
presence of Fe, Cr and Ni in the inter-crystallite region. The inset
displays an enlarged view close to $H=0$ indicating that demagnetization
effects are also observed for the $H$ // $ab$ and $H$ // $c$ orientations.
In a recent detailed study\cite{junod} the occurrence of Fe contamination
has already been identified, through measurements of MgB$_{2}$ samples made
from commercial powder supplied by a different company.

In view of the superimposed ferromagnetic signal in the magnetization
curves, we found to be not reliable to discuss the expected anisotropy in $%
J_{c}\propto \Delta M$, which could be determined using the Bean model\cite
{bean}. A rough estimate for both field orientations gives \ $J_{c}\simeq
10^{6}$ A/cm$^{2}$ at $H=1.5$ kOe and $T=5$ K (Fig. 3). \ This calculation
neglects the small influence of the ferromagnetic hysteresis and considers
the average crystallite geometry as described before. However, an anisotropy
between $J_{c}(H//c)$ and $J_{c}(H//ab)$\ should be expected. Indeed,
independently of the different regime of vortex pinning, $J_{c}$ \ is
predicted\cite{blatt} to be proportional to $\xi ^{2}$, leading to \ $%
J_{c}(H//c)$ / $J_{c}(H//ab)\approx (\xi _{ab}/\xi _{c})\approx
H_{c2}^{ab}/H_{c2}^{c}$.

A final cautionary observation has to be addressed to the possibility that
surface superconductivity could also be occurring for $H//ab$, since
coincidently the surface nucleation field is\cite{saint} $H_{c3}\simeq
1.7\,H_{c2}$. \ However, we have made several careful measurements of $M(H)$
and $\chi ^{\prime }$($H$), as well as $M(T)$ and $\chi ^{\prime }$($T$),
around the onset of transition, and no signature\cite{saint2} of a surface
nucleation field was found. This is consistent with the fact that our $%
H_{c2}^{ab}(T)$ line agrees with several reported $H_{c2}(T)$ lines measured
in polycrystalline samples\cite{budko,canfild,finemore,muller}, which
certainly did not comply the boundary condition\cite{saint,saint2} required
for surface nucleation in the $ab$ planes, i.e. $H//ab$.

In conclusion, we have prepared samples of aligned MgB$_{2}$ crystallites
that allowed, for the first time, the identification of an anisotropy for
the upper critical field given by $H_{c2}^{ab}/H_{c2}^{c}\simeq 1.73$,
implying an anisotropy of the coherence length $\xi _{ab}/\xi _{c}\simeq 1.73
$ and a mass anisotropy ratio $m_{ab}/m_{c}\simeq 0.3$. This could be
considered a mild anisotropy when compared to the values found for the
high-Tc materials ($m_{ab}/m_{c}\lesssim 0.04$). The influence of
contaminants is requiring further work, aimed at a more complete and
reliable characterization of the MgB$_{2}$ intrinsic properties. Naturally
the production of a good-sized single crystal of MgB$_{2}$ is also highly
desirable.

{\it Note added:} Since this manuscript was submitted two papers have
appeared\cite{khm,jim} showing results consistent with our anisotropy data.

\bigskip

We are very grateful to S. Gama for supplying some of the raw materials, to
I. Torriani and C. M. Giles for the X-ray diffractograms, and to D. Silva
(IG-Unicamp) for the SEM and microprobe analysis. \ This work is supported
by the Brazilian science agencies FAPESP (Funda\c{c}\~{a}o de Amparo a
Pesquisa do Estado de S\~{a}o Paulo) and CNPq (Conselho Nacional de
Desenvolvimento Cient\'{i}fico e Tecnol\'{o}gico).

\newpage \label{Refs.}

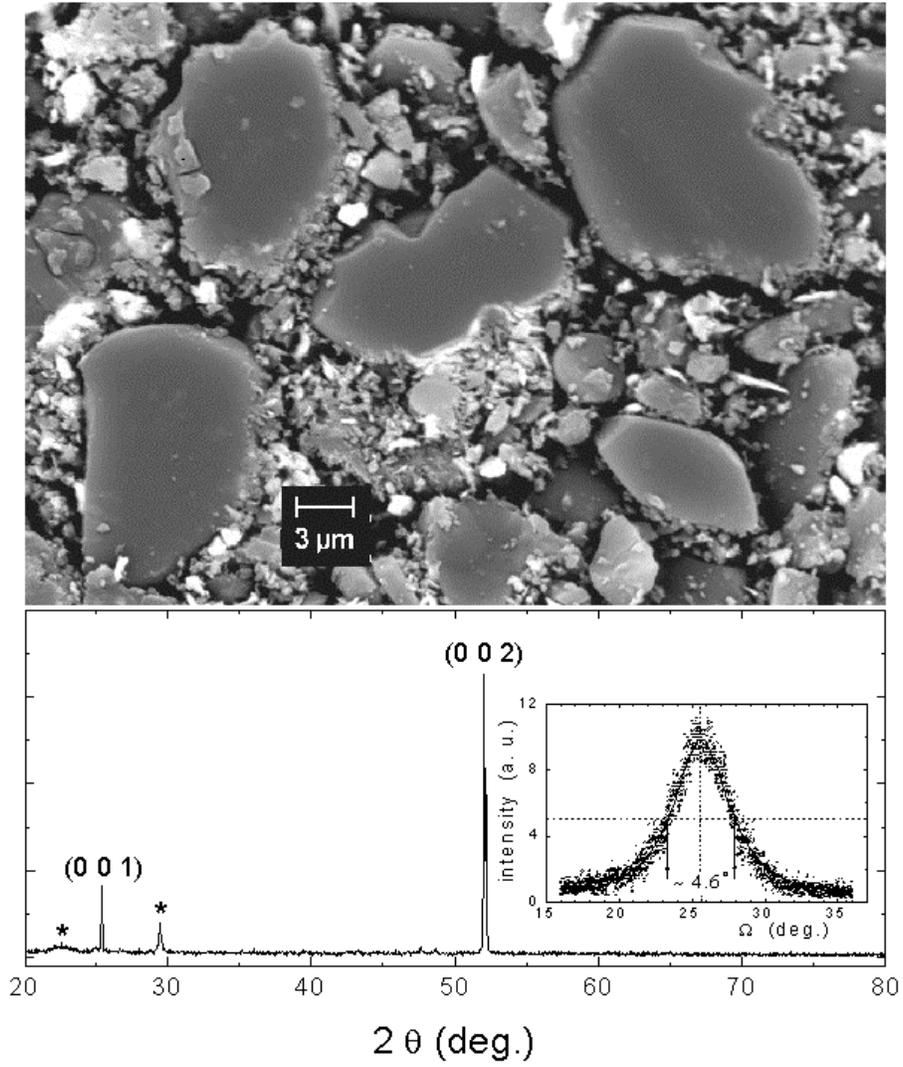
FIG. 1. Top: SEM picture showing the well aligned crystallites and
inter-crystallite material. Bottom: X-ray diffraction pattern showing only
the (001) and (002) peaks of MgB$_{2}$, plus two spurious peaks indexed as
SiO$_{2}$. Inset: rocking curve ($\omega $ scan) for the (002) peak, showing
an angular spread of about 4.6 degrees along the crystallites c axis.

\vspace{0.5cm}

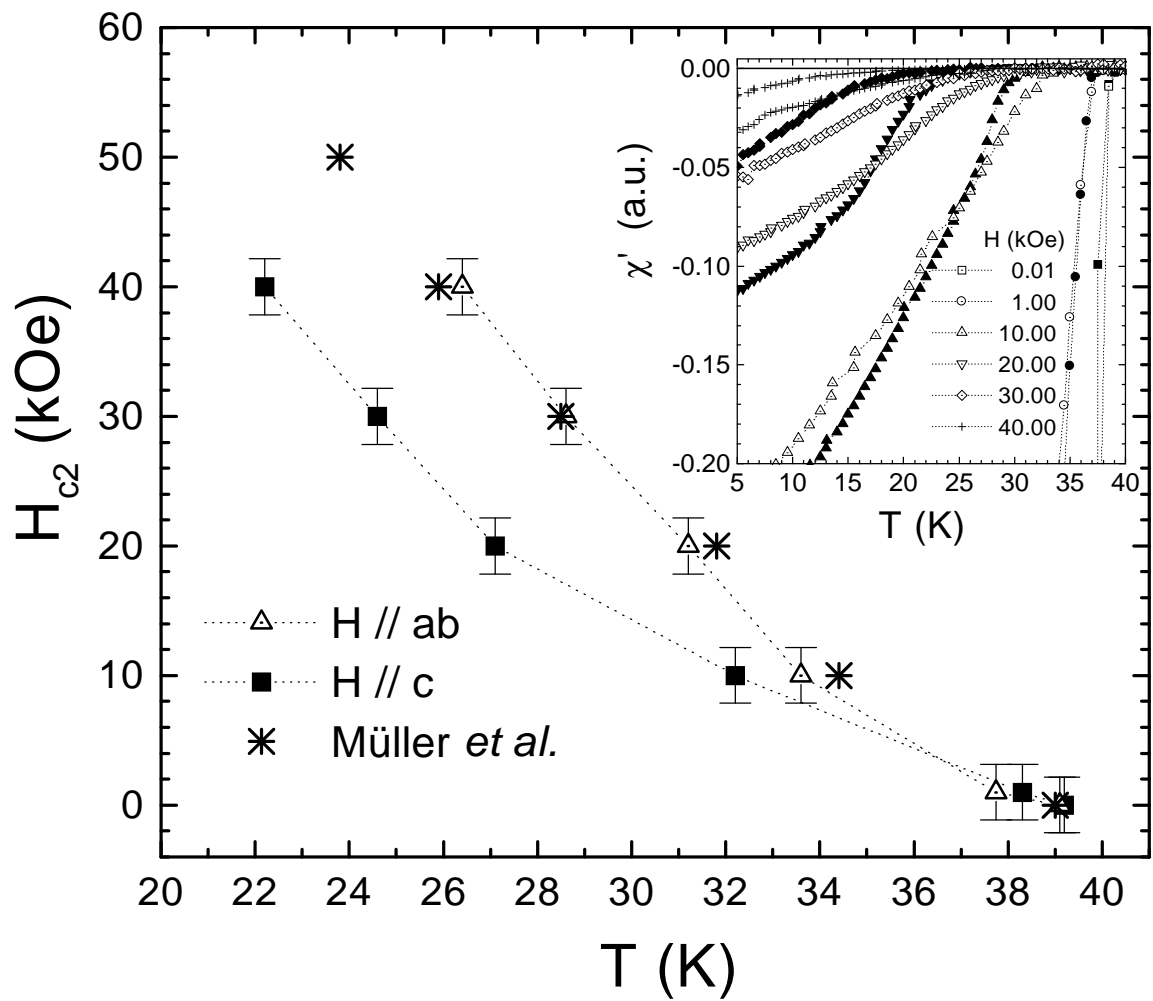
FIG. 2. Upper critical field $H_{c2}$ vs. Temperature phase diagram, for
both sample orientations. The stars represent the $H_{c2}$ vs. $T$ line from
Ref. 14. The inset shows the real component $\chi ^{\prime }$ of the ac
susceptibility vs. temperature, measured at several dc fields for both
orientations. Open symbols are for the $H//ab$ curves and solid symbols for $%
H//c$.

\vspace{0.5cm}

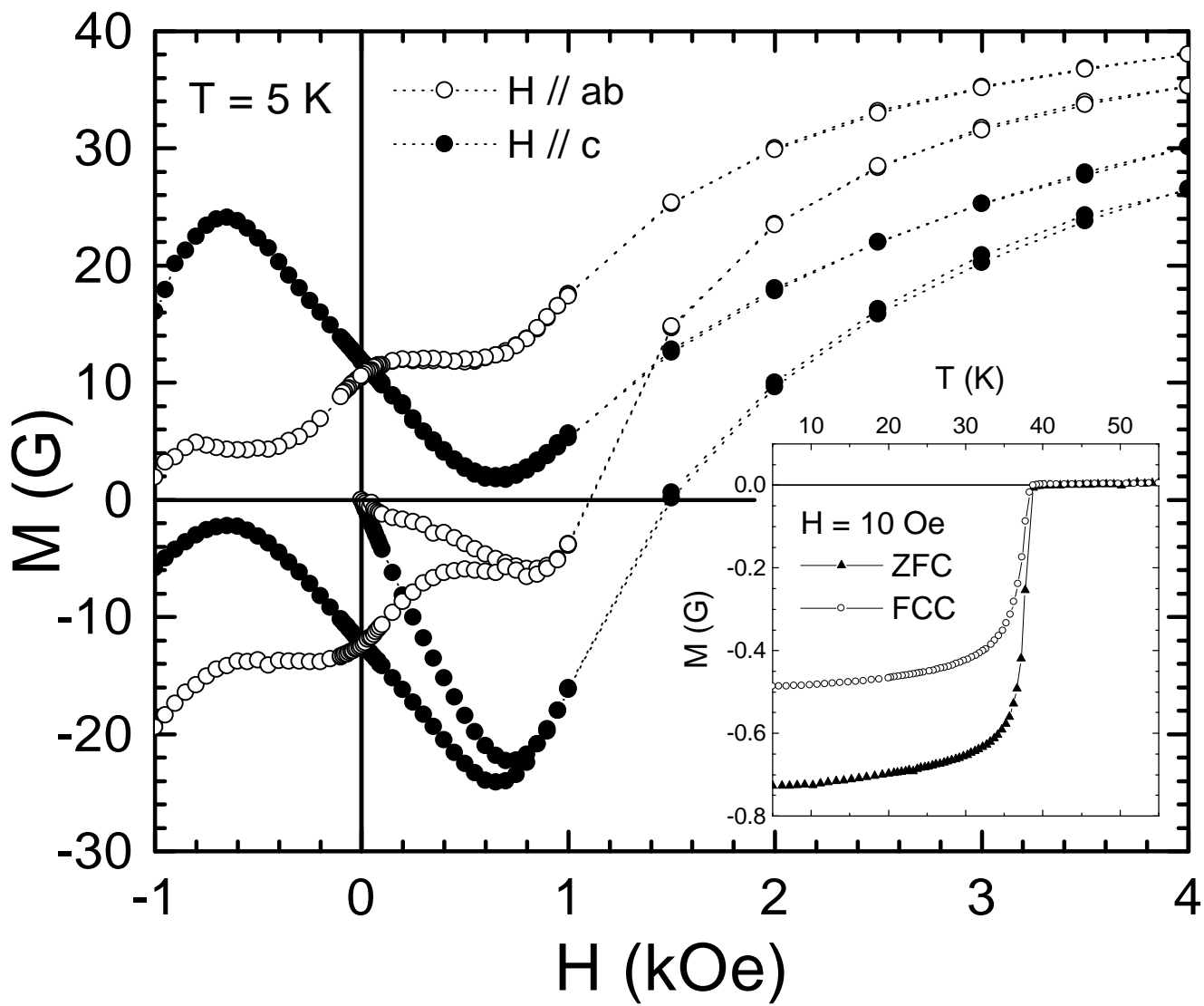
FIG. 3. Magnetization loops at 5 K for both sample orientations, showing a
superconducting hysteresis on a ferromagnetic background. The inset shows a
dc magnetization vs. temperature curve at 10 Oe, showing a sharp transition
at 39.2 K and $\sim 70$ \% recovery of diamagnetism for the FCC curve.

\vspace{0.5cm}

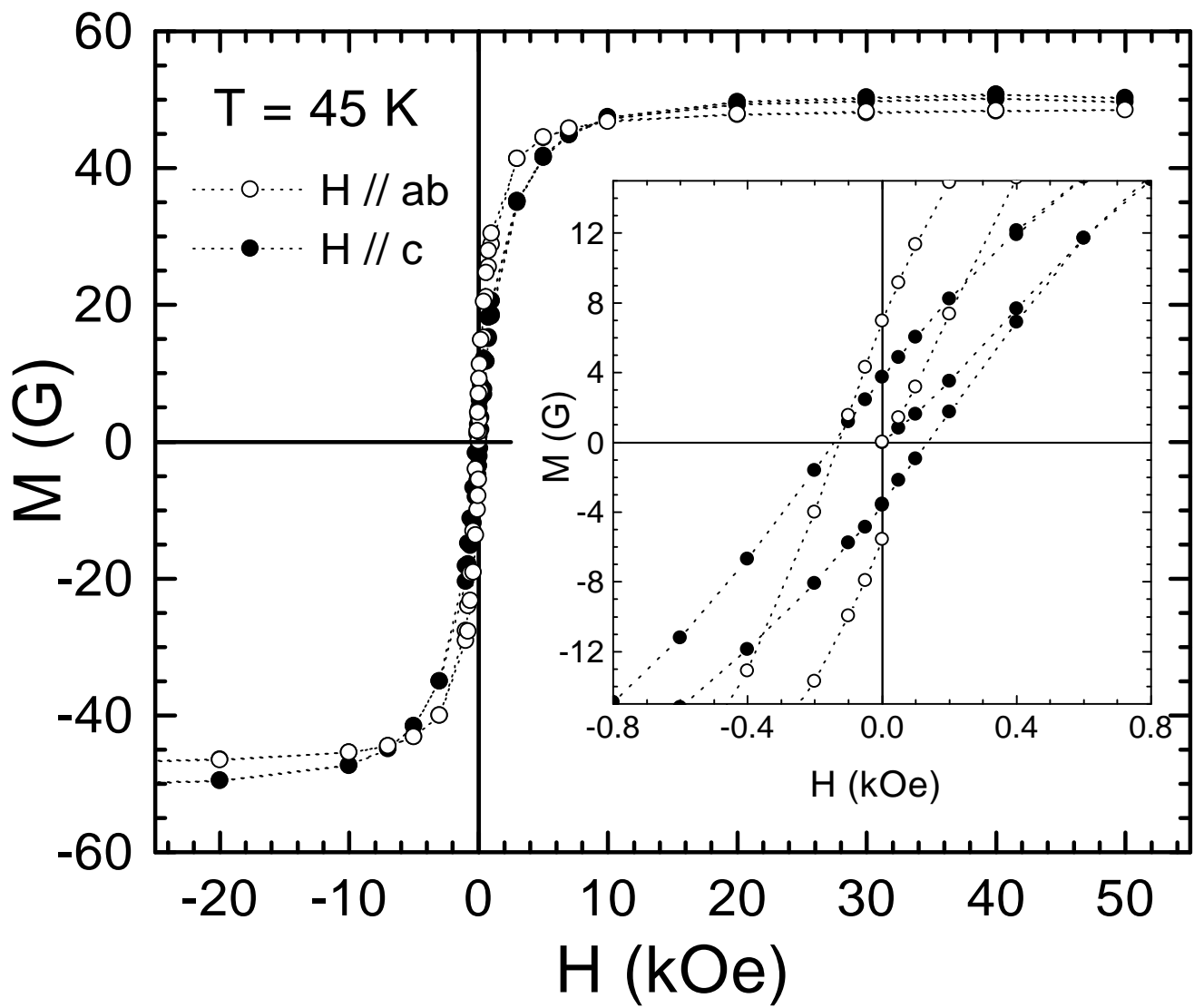
FIG. 4. Magnetization loops at 45 K (above $T_{c}$) for both sample
orientations, showing the ferromagnetic behavior of our sample. The inset
shows the hysteretic behavior at low fields.

\vspace{0.5cm}

\end{document}